# Simulating Isothermal Aging of Snow

Roman Vetter[1], Stephan Sigg[1], Herman M. Singer[1], Dirk Kadau[1], Hans J. Herrmann[1] and Martin Schneebeli[2]

[1] *IfB, Swiss Federal Institute of Technology ETH Zurich - 8093 Zürich, Switzerland*
[2] *WSL Institute for Snow and Avalanche Research SLF, Davos, Switzerland*

**Abstract.** - A Monte Carlo algorithm to simulate the isothermal recrystallization process of snow is presented. The snow metamorphism is approximated by two mass redistribution processes, surface diffusion and sublimation-deposition. The algorithm is justified and its parametrization is determined. The simulation results are compared to experimental data, in particular, the temporal evolution of the specific surface area and the ice thickness. We find that the two effects of surface diffusion and sublimation-deposition can accurately model many aspects of the isothermal metamorphism of snow. Furthermore, it is shown that sublimation-deposition is the dominant contribution for temperatures close to the melting point, whereas surface diffusion dominates at temperatures far below the melting point. A simple approximation of gravitational compaction is implemented to simulate density change.

**Introduction.** – The structural transformation of snow as a result of recrystallization has become an increasingly investigated field of research over the past few years. Isothermal metamorphism of snow was recently studied experimentally in Ref. [1], providing measurements of detailed microstructural properties. Most experiments were performed with in-vitro x-ray microtomography [2], while we used in-situ x-ray microtomography. This approach provides the possibility to directly analyze the evolution of the same snow crystals in space and time.

Isothermal metamorphism of snow is a sintering process between the ice grains forming the snow [3]. Various studies, although not always agreeing in every point [4–7], lead to assume that amongst the possible mass redistribution processes such as sublimation-deposition, grain boundary diffusion, plastic flow, surface diffusion, viscous flow and volume diffusion, only a few processes significantly contribute to the isothermal metamorphism of snow [1]. Surface diffusion and sublimation-deposition are believed to be the most dominant ones [8,9]. Isothermal metamorphism of snow has been stated to be similar to the evolution of bicontinous interfaces during coarsening [10].

Predictive simulations by means of numerical models of isothermal recrystallization of snow are hoped to lead to a deeper understanding of the governing processes of snow metamorphism and might become a helpful tool in applications such as avalanche forecasting. The current models [2, 11] use a macroscopic approach. Here we used a model which is based on a direct simulation by means of Monte Carlo processes, similar to the model of Ref. [12], but using effective ice particles instead of real molecules. The model presented in this paper implements the two dominant mass redistribution effects in isothermal metamorphism, for the effective ice particles. In order to directly compare and validate the simulation results with the experimental ones, important structural quantities, the specific surface area and the shape of the ice matrix, are measured and compared.

**Experiments.** – Experimental data has been obtained by Kaempfer and Schneebeli [1] from four samples of freshly-fallen snow, which was sieved at $-20°$C through a 1 mm sieve and filled in cylindrical sample holders with a diameter of 18.5 mm, which were sealed to avoid sublimation. The samples were kept for about 11 months in a temperature controlled environment at four different constant temperatures: $-54°$C, $-19.1°$C, $-8.3°$C and $-1.6°$C. Roughly once per month, the samples were i) weighed, ii) the snow height in the sample holders was determined, and iii) the samples were imaged with a mi-cro computer tomograph ($\mu$CT). During the $\mu$CT measurement, the samples were exposed to a temperature of $-15°$C for about two hours. The authors of Ref. [1] re-port that this temperature change did not affect the experiment as this period is considered to be short compared to the time scale of the processes involved in snow metamorphism. The $\mu$CT images were taken at a resolution of 10 $\mu$m per voxel. This resolution is chosen as it is lower than the typical roughness of snow [9]. The imaged snow region was about $5 \times 5 \times 5$ mm in size, which corre-sponds to a data block of $508 \times 508 \times 508$ voxels (e.g. see fig. 1). A marker was attached to this specific snow sam-ple to be able to visualize the temporal evolution of the same cutout as its position moves due to gravitational col-lapse. For direct comparison, in our simulations we used the same sample as in the experiment (fig. 1). The simu-lation parameters have been adapted as described in the section "Results". The pictures visually match well, both show a similar coarsening behavior.

**Modeling technique.** – Snow metamorphism is commonly modeled by macroscopic approaches using surface growth laws, e.g. the curvature driven model presented by Flin et al [2], or using phase field models, such as the recently presented temperature driven model of Kaempfer and Plapp [11]. A brief review on models for snow metamorphism can be found in Ref. [11]. Contrarily to studying the details of crystal growth on a microscopic scale the sublimation-deposition and diffusion processes of individual atoms are modeled, e.g. in Ref. [12].

Here, we will use a similar way of determining the sticking and sublimation probabilities for the sublimation-deposition process as well as the jump rate for diffusion as in crystal growth [12]. However, instead of considering individual atoms as commonly in crystal growth, we use "effective ice particles". For this reason we have to determine the parameters in the model by adjusting to experimental data as no first principle "microscopic" parameters exist for these coarse grained particles. Therefore, we use the same size and shape as in the experiments ($10\mu$m cubes, see previous section). These effective particles contain the complex interactions of microscopic properties.

The term *sublimation-deposition*, within this context, refers to the three-step process of single particles vaporizing from the snow-air interface into the air, vapor transport within the air, and eventually, single vapor particles depositing again on the surface. The steps of sublimation and deposition can easily be physically modeled. Important steps of the derivation will be given later in this section. Monte Carlo processes are based on proba-bilistic rules. Xiao et al. [12] derived the probabilities of growth units to stick on crystal surfaces for a two-dimensional triangular lattice, based on the surface-free-energy minimization principle, using nearest and second-nearest neighbor interactions. In this section we derive in an analogous manner the probabilities of an effective ice particle to transform from the solid state on the surface into vapor and vice versa for the lattice given by the cubic voxel structure, and taking interactions to all 26 neighbors in the local $3^3$ cube into account, i.e. the 3D Moore neighborhood [13, 14]. In the following this definition of neighbors is used.

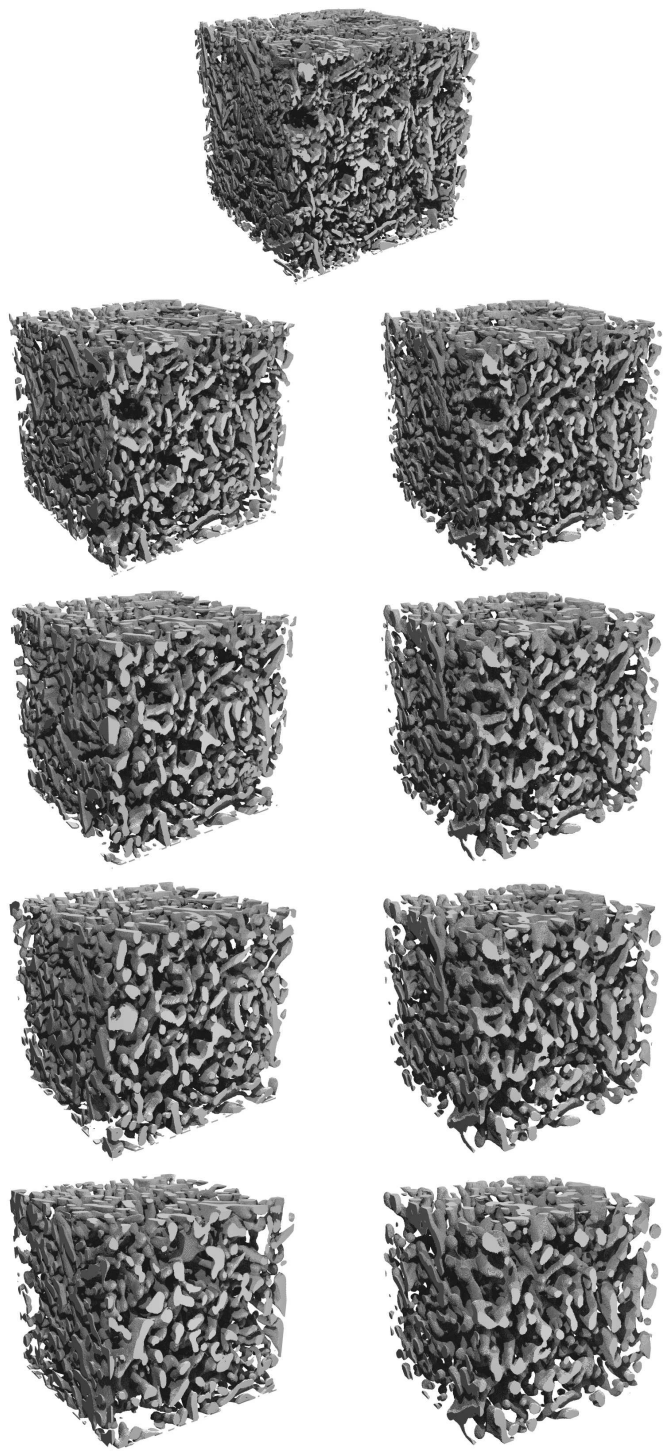

Figure 1: Snapshots of real samples used in experiments (left) compared to the simulated configuration (right) for a temperature of $-8°$C (top to bottom: initial configuration, after 2 weeks, 5 weeks, 7 weeks, and 10 weeks). As starting configuration for the simulation the experimental data has been used. Here, in the experiments the same cutout of the snow sample is shown. Due to unequal settling in the experiment, the sample shows a slight "drift" to the right, whereas the simulation box is fixed.

A general expression for the sublimation rate $K_i^-$ can be written as [12]

$$K_i^- = \nu \, exp\left(-\frac{E_i}{k_B T}\right) = \nu \, exp\left(-\frac{\phi n_i}{k_B T}\right) \quad (1)$$

where $\nu$ is a vibration factor [12] or frequency factor which usually also depends on temperature [15], $k_B T$ the thermal energy and $E_i = \phi n_i$ the total binding energy of particle $i$, with $\phi$ the interaction energy of one particle with one neighbor, assumed to be the same for all neighbors, and $n_i \in \{0, ..., 26\}$ the number of occupied neighbors of site $i$.

Similarly the attachment rate $K^+$, at which particles impinge onto the surface, can be written as [12]

$$K^+ = K_{eq} exp\left(\frac{\Delta \mu}{k_B T}\right), \quad (2)$$

where $K_{eq}$ is the equilibrium value of $K^+$ and also temperature dependent ($K_{eq}(T)$), $\Delta \mu = \frac{\Delta \overline{G}}{N_A}$ the average chemical potential difference per vapor particle with $\Delta \overline{G}/N_A = RT \ln \frac{\overline{p}}{p_{eq}}/N_A$ the average Gibbs-free-energy at constant temperature in supersaturated vapor with respect to the bulk average [16] per mole.

Considering the equilibrium condition $\overline{p} = p_{eq}$, $n_i^{eq} = \frac{3^3-1}{2} = 13$, $K_{\overline{i}}^- = K^+$, we can eliminate $K_{eq}$:

$$K_{eq} = \nu \frac{exp\left(-\frac{E_i}{k_B T}\right)}{exp\left(\frac{\Delta \mu}{k_B T}\right)} = \nu \, exp\left(-\frac{13\phi}{k_B T}\right). \quad (3)$$

For $\gamma := exp\left(\frac{\Delta \mu}{k_B T}\right) = \frac{\overline{p}}{p_{eq}}$ and $\beta := exp\left(-\frac{\phi}{k_B T}\right)$ we thus find the probability, that particle $i$ sticks to the surface to be

$$P_i^{stick} = \frac{K^+}{K_i^- + K^+} = \frac{\gamma \beta^{13-n_i}}{1 + \gamma \beta^{13-n_i}} \quad (4)$$

and accordingly, the probability to evaporate off the surface

$$P_i^{leave} = 1 - P_i^{stick} = \frac{1}{1 + \gamma \beta^{13-n_i}}. \quad (5)$$

These probabilities directly depend on the number of neighbors $n_i$ and the current vapor pressure ratio $\gamma = \frac{\overline{p}}{p_{eq}}$. In the simulations, presented here, the initial value $\gamma_0$ will be set to unity, i.e. the initial average pressure is the equilibrium pressure. For an isothermal process $\beta$ is constant. In a similar way, the initial water vapor volume fraction $f_V$, which (together with $\gamma_0$) determines the initial number of vaporous ice particles, is set to the equilibrium one (using the ideal gas law). The calculated values are also listed in table 1.

Vapor transport is, as a first approximation, modeled by a random walk. As the connecting component between sublimation and deposition, the vapor transport must ensure that in equilibrium sublimation and deposition rates are identical. This is achieved by performing the random walk only for one single vapor particle at a time until it impinges onto a surface.

In practice, at each time step, for as many times as specified by an evaporation rate $K_V$ a particle from the surface is evaporated with probability $P_i^{leave}$, i.e. while keeping its position the particle is added to the vapor container consisting of a list of vapor particles and their positions. Then, a randomly chosen particle of the vapor container performs a random walk until it reaches the ice surface. It sticks at the surface with probability $P_i^{stick}$, i.e. deleted from the vapor container and added to the ice matrix. Otherwise it stays within the container.

*Surface diffusion* describes the diffusive motion of particles on the surface of a solid material. It changes the microstructure of the surface and results in smoothing rough surface regions. The surface diffusion rate depends on the strength of the particle-particle interaction. A practical way to approximate it is to consider the interactions to all neighbors, i.e. the number of neighbors multiplied with a constant interaction energy. As for the sublimation-deposition process, we need to take into account 26 possible directions for each move, each of which is chosen with the same probability if the new site is unoccupied and the particle remains on the surface. The jump rate $K_{i \to j}$ from site $i$ to site $j$ can be written in the following way [12]:

$$K_{i \to j} = \nu \, exp\left(-\frac{\Delta E_{ji}}{k_B T}\right) \quad (6)$$

The activation energy $\Delta E_{ji}$ is the change of energy caused by this move, $\Delta E_{ji} \approx \phi(n_i - n_j)$ where $\phi$ denotes the interaction energy between a pair of neighboring particles, $n_i$ and $n_j$ the numbers of neighbors of site $i$ and $j$, respectively, and $\nu$ a vibration or frequency pre-factor (see above). In practice, at each time step, for as many times as specified by a diffusion rate $K_D$, a random surface voxel $i$ is selected for which the moving direction is chosen randomly with two constraints: there must not already exist a voxel $j$ at the target site of the move and the voxel must remain on the surface. The second constraint ensures that no voxel leaves the surface during the surface diffusion process. If both constraints are fulfilled, the move is executed with the Metropolis probability $P_{i \to j} = \min \{1, exp\left(-\phi(n_i - n_j)/k_B T\right)\}$ which is essentially equation (6) but with the vibration pre-factor included into the diffusion rate. Note that for fresh snow samples the surface structure is such that for most cases a diffusion process leads to a lower energy and is therefore accepted. In these cases the speed of the coarsening process due to surface diffusion is determined only by the diffusion rate $K_D$.

Summarizing, the most important parameters are the interaction energy $\phi$, a temperature independent property of the effective ice particles, and the evaporation and dif-

R. Vetter, S. Sigg, H.M. Singer, D. Kadau, H. J. Herrmann, M. Schneebeli

fusion rates $K_V$ and $K_D$ which are both temperature dependent as the vibration factor depends on $T$. To be used in the simulations these parameters have to be determined by comparison to the experimental results which will be described in the next section. First, we will present results neglecting the compaction of the snow due to gravity. Later in this paper, we introduce the effect of *gravity* by using a simple compaction scheme.

**Results.** – A simple qualitative criterion for the correctness of the simulation results is given by visual inspection. Since Kaempfer and Schneebeli [1] published a series of snapshots from the recorded $\mu$CT images, we can simply check if the visual outcome of the program fits the experimental observation, i.e. matching morphological measures such as the specific surface area or the ice thickness (see below) judged "by eye". Figure 1 shows one example of a short term series (10 weeks) for a system of $508 \times 508 \times 508$ voxels corresponding to 5 mm $\times$ 5 mm $\times$ 5 mm in the experiments to illustrate that visual matching could be achieved reasonably well. Note that considering that the volume fractions of the experimental sample is around 30% the number of effective ice particles of this simulation is about 40 million. For more detailed comparison we used the original images by Kaempfer and Schneebeli for the long term behavior (about 11 months), i.e. system sizes of $192 \times 192 \times 192$ voxels corresponding to 2 mm $\times$ 2 mm $\times$ 2 mm in the experiments. Considering that the volume fractions of the experimental samples are around 30% the number of effective ice particles is over 2 million. Table 1 shows the simulation parameters used to match the experiments at the temperatures indicated [1]. Note, that the parameters for $k_B T/\phi$ are all using the same value for $\phi$, which is the parameter we had to determine.

Table 1: The four simulation settings used to reproduce the experiment. The parameters $\phi$, $K_V$ and $K_D$ used in the simulations (single runs) are determined such that the images match the experiments well. $f_V$ (in brackets) is calculated assuming the ideal gas law for the effective ice particles.

| Setting, T | $k_B T/\phi$ | $K_D$ | $K_V$ | $(f_V)$ |
|---|---|---|---|---|
| 1, -54°C | 0.80 | 5 | 1 | (0.0001) |
| 2, -19°C | 0.93 | 19 | 17 | (0.0022) |
| 3, -8°C | 0.97 | 26 | 40 | (0.0043) |
| 4, -2°C | 0.99 | 32 | 88 | (0.0061) |

Note, that rescaling the evaporation and diffusion rates (both multiplied by the same factor) mainly leads to an effective rescaling of the simulation time, i.e. only their ratio is important and in principle they determine the time scale in the simulations. At low temperatures compaction is practically negligible as found in the experiments [1]. The higher the temperature, the larger the density discrepancy between experiment and simulation output, where the compaction is not considered.

Similarly to the experiments we analyze the specific surface area as a function of time. According to Ref. [17] the

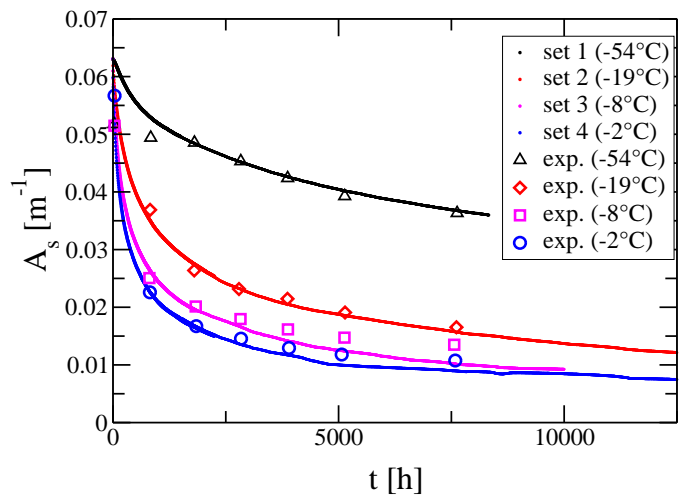

Figure 2: Specific surface area $A_s$ depending on time for the simulations and experiments. The simulation steps are scaled to real time such that they match the experiments best.

specific surface area $A_s$ obeys the law

$$A_s(t) = A_{s,0} \left( \frac{\tau}{\tau + t} \right)^{1/n} \qquad (7)$$

with the initial specific surface area $A_{S,0}$, growth exponent $n$ and a parameter $\tau$, which determines the time scaling. Our simulation results strongly confirm this relationship which is a non-trivial experimental measurement. In all simulations the squared correlation coefficient $R^2$ was larger than 0.999. Figure 2 shows the evolution of the specific surface area $A_s$ for the four parameter sets (tab. 1). Here we deliberately chose to present results for single runs to directly compare to experiments. Due to the large number of modeled ice particles still a rela-tive good average could be achieved. Superimposed are the experimental results. Note, that the simulation steps are rescaled to real time, such that they match the experiments. For high temperatures (-2°C, -8°C) the long term behavior differs, which is expected due to neglecting the density increase due to gravity in the model for the simulations presented above. For lower temperatures the long term behavior matches well, as in that case the sample density does not increase much. For set 1 (-54°C) the initial volume fractions were different, so that only the long term behavior could be reproduced by the simulations.

The fit parameters according to eq. (7) are compared in table 2. The simulations show a clear trend indicating that the exponent $n$ decreases with increasing temperature. However, for the experiments it is more difficult to determine this exponent very accurately as the number of measured points is limited, leading to differences between the exponents of simulation and experiments. For very low temperatures the simulation relaxes too fast in the beginning, compared to the experiment. This can be explained as the simulation uses Metropolis rates in accor-dance with the hopping probability for atoms on a crystal surface. Thus, all diffusion processes lowering the energy are accepted, which seems to be unrealistic for the effective ice particles.

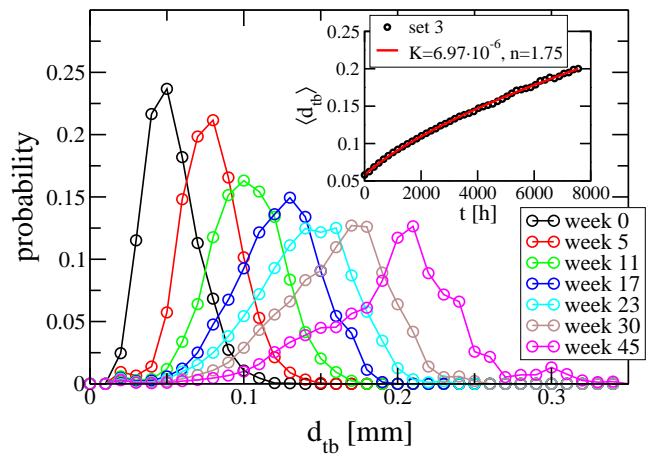

Figure 3: Distribution of the ice thickness $d_{\text{tb}}$ for several snapshots at different times for the simulations. The inset shows the corresponding time dependence of the average ice thickness $\langle d_{\text{tb}} \rangle$, including a fit according to eq. (8).

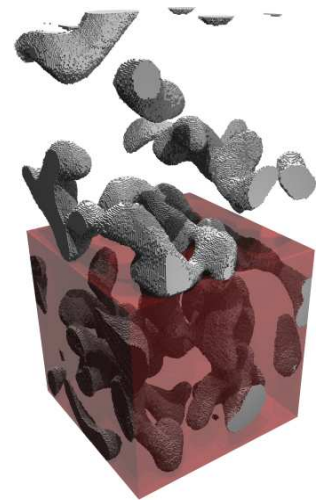

Figure 4: Illustration of the density measurement in the simulation. By using the simplified gravity rule only the lower part of the system could be used for measuring the density.

Table 2: Fit parameters obtained for the four simulation settings (including error) and the corresponding experiments. The simulation steps are rescaled to real time as in fig. 2.

| Setting, T | $n_{\text{exp}}$ | $n_{\text{sim}}$ | $\tau_{\text{exp}}$ | $\tau_{\text{sim}}$ |
|---|---|---|---|---|
| 1, -54°C | 0.8 | 5.89±0.05 | 25000h | (470±5)h |
| 2, -19°C | 2.20 | 2.93±0.01 | 652.1h | (175±1.5)h |
| 3, -8°C | 3.55 | 2.60±0.01 | 69.97h | (93±0.7)h |
| 4, -2°C | 3.00 | 2.44±0.01 | 55.53h | (73±0.5)h |

Considering that the simulation parameters were only determined using single runs for each parameter set and by visual matching of the figures compared to the experiments as well as adjusting the time scale in accordance to the specific surface area, the simulations match the experiments relatively well. In future investigations better statistics and a refined parameter determination by matching morphological measures as the specific surface area and ice thickness quantitatively are necessary. For that purpose also more experimental data would be needed (see also outlook). Additionally, investigating the temperature dependence, e.g. of the parameters $K_V$ and $K_D$ (i.e. their ratio, see above), is crucial to predict the behavior of snow samples for all possible temperatures without the need of new eperimental measurements.

The thickness of the ice matrix is another important quantity which has been measured in the experiments. Here, we will use the thickness definition introduced by Hildebrand and Rüegsegger [18] which can be used for arbitrary three-dimensional structures. The local thickness of a porous volume at point $P$ inside the structure is defined as the diameter of the sphere with maximal radius which lies completely inside the structure and contains $P$. This definition gives a somewhat intuitive measure of the local thickness of a matrix trabecula. Each spatial point of the structure (and in particular each snow voxel) is assigned a positive distance describing the structure thickness at that point. By measuring these thicknesses for each voxel of an ice matrix (at given time) one gets a thickness distribution, and can determine the average thickness.

Legagneux and Dominé [17] proposed the following coarsening law for the average ice thickness $\langle d_{\text{tb}} \rangle$ for snow:

$$\langle d_{\text{tb}} \rangle (t) = (\langle d_{\text{tb},0} \rangle^n + Kt)^{\frac{1}{n}} \quad (8)$$

with growth rate $K$, growth exponent $n$ and initial ice thickness $\langle d_{\text{tb},0} \rangle$. Our simulations agree with this function with very high precision (inset in fig. 3), thus confirming the validity of eq. (8). The ice thickness distribution has been studied by Kaempfer and Schneebeli [1]. The main feature observed in the experiment is a broadening of the distribution. As depicted in figure 3 for setting 3 the same behavior is found for the simulations.

A difference between the experimental $\mu$CT image measurements and our simulations discussed in the previous section is the absence of density increase. Therefore, we introduced a simple gravity effect. As soon as a grain (piece of snow) is not connected with the bulk anymore it falls down to the point where at least one point on the surface is supported by another piece of snow from below. This simplified model of gravity simulates the compaction of snow. However, mechanical behavior like rotation, translation and breaking are neglected. During this process, when using a fixed size for the simulation box, the upper part of the system dilutes successively. Therefore, we only measure the density in the lower part of the system (here the lower half), as illustrated in fig. 4. Therefore, we used elongated systems of size of $192 \times 192 \times 384$ for the following measurements.

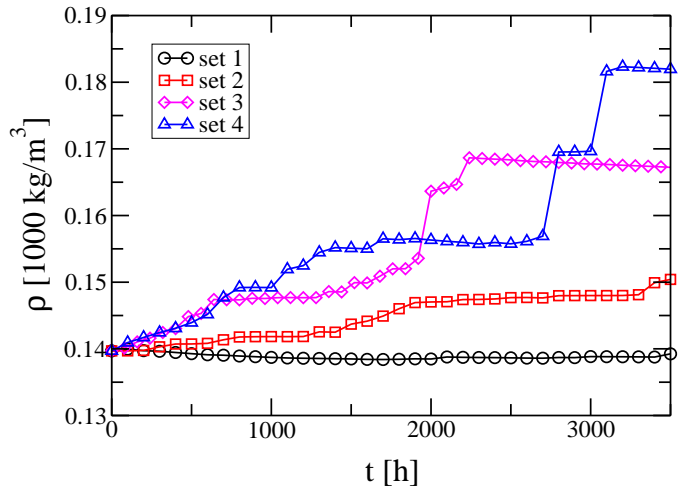

Figure 5: Time evolution of the density in the simulation using a simple model for gravity. The simulation steps are rescaled to real time as in fig. 2.

The density measurements as depicted in figure 5 features a step form as a consequence of the fact that not at every time step unconnected clusters are found and moved downwards. This behavior is also observed in the experiments [1], but it is much less pronounced due to the larger sample size. The decrease in density between the jumps can be explained by the fact that to measure the density in the simulations only the lower half of the system is used (see above, fig. 4). Therefore, effective ice particles sublimated from the ice matrix in the lower half will move to the upper half where more free space is available. On average, the density increase measured in the simulation is much lower than in the experiments [1] for all four settings. This is expected as we use a strongly simplified model for gravity, only considering the most important process, the "falling" of the grains, but neglecting processes like sliding and rolling of grains, or breakage due to the load of the snow above, which certainly play a role in real snow samples.

**Conclusion & Outlook.** – A new simulation technique using effective ice particle for the aging of snow is presented. We show that our Monte Carlo simulation incorporating the two basic processes of snow metamorphism, sublimation-deposition and surface diffusion, can adequately model the experimentally observed behavior of isothermal snow metamorphism. In accordance with Libbrecht [19], a transition to more surface diffusion dominated coarsening takes place at lower homologous temperatures. In our simulations we could verify previously suggested laws originating from classical sintering theory [20] for the time evolution of the specific surface area and average ice thickness. While classical sintering theory assumes that an integer exponent $n$ uniquely describes the process, our exponents $n$ seems to vary continuously. We ascribe this to the fact that the small range in homologous temperature where vapor and surface diffusion interact leads to a transition from one process to the other. Until now, it has been unclear if the non-integer exponents deduced from experiments were an artifact or real. The simulation supports the view that these non-integer exponents are a real phenomenon of isothermally sintering snow. To be able to use the simulations as a predictive tool for snow metamorphism, a systematic parameter determination is needed. This will be an issue for future studies as a better resolution in the experiments would be needed to achieve more accurate parameters. The results of the simple gravity model are realistic for very small strain rates as those in the experiment. High strain rates will create a much more complex behavior.

∗ ∗ ∗

We thank M. Heggli, B. Pinzer and H. Löwe for support in handling the data and suggestions to the manuscript.